\documentclass{article}
\usepackage{amsmath}
\usepackage{graphicx}
\def\p{\partial}

\def\de{\delta}

\def\De{\Delta}
\def\ov{\overline}

\def\e{\eta}
\def\et{\eta}

\def\Om{\Omega}

\def\b{\beta}

\def\a{\alpha}

\def\pdellx'{\frac{\partial}{\partial x'}}
\def\pdellw'{\frac{\partial}{\partial w'}}

\newcommand{\be}{\begin{equation}}
\newcommand{\ee}{\end{equation}}
\def\bed{\begin{displaymath}}
\def\eed{\end{displaymath}}
\def\bea{\begin{eqnarray}}
\def\eea{\end{eqncrray}}
\def\[{$$}
\def\]{$$}
\begin{document}
\title{Dynamic Models for the Beginning, Hubble Law and the Future of the Universe  Based on Strong Cosmological Principle and Yang-Mills Gravity}
\author{
Jong-Ping Hsu\\
 Department of Physics, 
 University of Massachusetts Dartmouth,\\ 
North Dartmouth, MA 02747, USA\\
Leonardo Hsu\\
Department of Chemistry and Physics, \ \   Santa Rosa Junior College,\\
Santa Rosa, CA 95401, USA}

\maketitle
{\small We discuss highly simplified dynamic models for the beginning, expansion and future of the universe based on the strong cosmological principle and Yang-Mills gravity in flat space-time.   We derive a relativistic Okubo equation of motion for galaxies with a time-dependent effective metric tensor $G_{\mu\nu}(t)$.  The strong cosmological principle states that $G_{\mu\nu}(t)=\e_{\mu\nu} A^2(t)$ for $t \ge 0$.  In a model (HHK) with Yang-Mills gravity in the super-macroscopic limit, one has $A(t)= a_o t^{1/2}$, which leads to the initial mass run away velocity $\dot{r}(0)=c$, associated with $r(0)=r_o>0$.  Thus, the Okubo equation of motion for galaxies predicts a `detonation' at the beginning of the universe.  The Okubo equation also implies $r(\infty) \to \infty$, $\dot{r}(\infty) \to 0$ with zero redshift for the future of the universe.   In addition, the Okubo equation leads to the usual Hubble's law $\dot{r}(t) \approx  H(t) r(t)$, where $H(t)=\dot{A}(t)/A(t)$ in non-relativistic approximation.  We also discuss a model with a strict Hubble linear relation $\dot{r}(t) \approx  const.\times r(t)$ for all time. This model gives a silent beginning of the universe: $\dot{r}(t)=0, \ \ddot{r}(t)\to\infty$ as $t \to 0$; and final radius $r(t) \to \infty,$ final velocity, $\dot{r}(t) \to c$, $\ddot{r}(t)\to 0$ as $t \to \infty$.  In all models with the strong cosmological principle in flat space-time, Hubble's recession velocities are predicted to have a maximum, i.e.,  the speed of light, as measured in an inertial frame. 
  


\section{The macroscopic and super-macroscopic pictures of the universe}

At macroscopic scales, the laws of motion for classical objects and light rays  are the geometric-optics limit of wave equations in quantum field theory such as quantum Yang-Mills gravity and QED.\cite{1,2,3}  One usually expects that  physics at the macroscopic scales could be applied to cosmological or super-macroscopic scales.  However, this turns out to be not the case. 

There is a basic difference between the geometries of space-time in usual  macroscopic scales and that of the universe as a whole.  This difference shows up in the recession motion of distant galaxies or in Hubble's law. 
Roughly speaking, the metric tensor of space-time for the motion of usual macroscopic objects such as planets or light rays are associated with (or approximated by) a static function $G_{\mu\nu}(x,y,z)$.   In Yang-Mills gravity\cite{4}, the macroscopic action $S_{ma}$ for a classical object with mass m is postulated to be 
\be
S_{ma}=-m\int ds, \ \ \ \   ds^2= G_{\mu\nu}(x,y,z)dx^\mu dx^\nu, \ \ \ \ \ \   c=\hbar=1,
\ee
 where $G_{\mu\nu}$ depends on the $T_4$ tensor field and is an `effective metric tensor'  in flat space-time.  The action (1) has little to do with curved space-time, although it is formally the same as that in general relativity.  Yang-Mills gravity reveals that
  the apparent curvature of space-time appears to be simply a manifestation of the flat space-time translational gauge symmetry for the motion of quantum particles in the geometric-optics limit.\cite{5}
 
  In contrast, the metric tensor of space-time of the universe as a whole is closely related to the cosmological principle of homogeneity and isotropy. In this case, the metric tensor does not depend on space coordinates and its non-diagonal components vanish.  Thus, it could be approximated by a time-dependent function $G_{\mu\nu}(t)$ in the super-macroscopic limit of Yang-Mills gravity.  And the Okubo (or super-macroscopic) action for a distant galaxy with mass M is assumed to be 
 \be
 S_{Ok}= -M \int ds,      \ \ \ \ \    ds^2=G_{\mu\nu}(t)dx^\mu dx^\nu,
 \ee
 where a galaxy is idealized as a mass point.  Such a time-dependent property of  $G_{\mu\nu}(t)$ could be related to the homogeneity of the universe.
Within quantum Yang-Mills gravity formulated in inertial frames, the metric tensors in $S_{ma}$ and $S_{Ok}$ are obtained from the geometric-optics limits of the quantum wave functions.\footnote{One obtains Hamilton-Jacobi type equation of motion for classical objects in the geometric-optics limit, and then the equation can be derived from a corresponding action.}

Quantum Yang-Mills gravity is a new gauge symmetry theory based on the local space-time translation ($T_4$) symmetry in flat space-time.\cite{3}   It gives an elegant explanation as to why the gravitational force is always  attractive between particle-particle and between particle-antiparticle.
This property of attractive gravitational force is embedded in the coupling between the gravitational tensor field $\phi_{\mu\nu}$ and the matter fermion field $\psi$ at the quantum level.  Let us consider the  gravitational ($T_4$) tensor field $\phi_{\mu\nu}(x)$ and the electromagnetic potential field $A_\mu(x)$ in the gauge covariant derivative of the electron wave function.  The complex conjugate of the gauge covariant derivative is associated with the positron wave function:

\be
 \p_\mu -ig\phi_\mu^\nu p_\nu -ie A_\mu +.... = \p_\mu +g\phi_\mu^\nu \p_\nu -ie A_\mu +....   \ \ \ \ for \ \psi(e^-),
\ee
\be
 (\p_\mu -ig\phi_\mu^\nu p_\nu -ie A_\mu +....)^* = \p_\mu +g\phi_\mu^\nu \p_\nu + ie A_\mu +....  \ \ \ \  for \ \ \psi(e^+).
\ee
The electric force between two charged particles is due to the exchange of a virtual photon in QED.  The key properties of the electric force $F_e(e^-,e^-)$ (i.e., between electron and electron) and the force $F_e(e^-, e^+)$ (i.e., between electron and positron) are given by the third terms in (3) and in (4), i.e.,
\be
F_e(e^-,e^-): \ \ (-ie)\times (-ie)= - e^2, \ \ \ \ \   repulsive, 
\ee
\be
F_e(e^-, e^+): \ \  (-ie)\times (ie)=+e^2,  \ \ \ \ \ \   attractive.
\ee
The force $F_e(e^+,e^+)$ can be verified to be the same as $F_e(e^-,e^-)$. Thus, the experimentally established attractive and repulsive electric forces are due to the presence of $i$ in the electromagnetic $U_1$ gauge covariant derivative, which dictates the electromagnetic coupling in QED. 

 In contrast, the Yang-Mills gravitational force $F_{YMg}(e^-, e^-)$ (i.e., between electron and electron) and the force $F_{YMg}(e^-, e^+)$ (i.e., between electron and positron) are respectively given by the second terms in (3) and in (4).   Because the gravitation coupling terms in (3) and (4) do not involve $i$, we have attractive rather than repulsive gravitational force,
\be
F_{YMg}(e^-, e^-) : \ \ \ (g) \times (g)= +g^2, \ \ \ \ \   attractive,  
\ee
\be
F_{YMg}(e^-, e^+) : \ \ \  (g) \times (g) =+g^2, \ \ \ \ \   attractive.
\ee
One can also verify that the Yang-Mills gravitational force $F_{YMg}(e^+, e^+)$ is the same as $F_{YMg}(e^-, e^-)$.  Furthermore, the gravitational coupling constant $g$ in (3) and (4) has the dimension of length (in natural units), in contrast to all other coupling constants of fields associated with internal gauge groups, so that $g^2$ is naturally related to Newtonian gravitational constant $G$ by $g^2=8\pi G$.\cite{7,3}

 The theory of Yang-Mills gravity is consistent with classical experiments such as the perihelion shift of Mercury, deflections of light rays by the sun, redshifts and gravitational quadrupole radiations.\cite{2,3}  It has been quantized to derive the gravitational Feynman-Dyson rules for Feynman diagrams,\cite{4} and it has brought gravity back to the arena of gauge field theory and quantum mechanics.   
  
  In the macroscopic limit, we derive a relativistic Hamilton-Jacobi type equations involving an effective metric tensor $G^{\mu\nu}(x)$ for classical objects and light rays.  These equations derived in flat space-time turn out to be formally the same those in the approach of Einstein-Grossmann.  The key difference is that $G^{\mu\nu}(x)$ in YM gravity is actually a function of the spin 2 tensor gauge fields in flat space-time and has little to do with the curved space-time.\cite{6}.  In this connection, we arrive at a new viewpoint of gravity:
  
  Yang-Mills gravity reveals that
  the apparent curvature of space-time appears to be simply a manifestation of the flat space-time translational gauge symmetry for the motion of quantum particles in the geometric-optics limit.\cite{5} 

The basic framework of particle-cosmology, including quantum Yang-Mills gravity, should be based on the four-dimensional space-time with Lorentz-Poincar\'e invariance in inertial frames, for simplicity.  Such a framework can be generalized to include non-inertial frames, based on the principle of limiting continuation of physical laws.\cite{10}  

We explore the implications of particle-cosmology based on a `strong cosmological principle' and Yang-Mills gravity.  The solutions of the highly simplified cosmic Okubo equation for galaxies at all times give a qualitative picture of the total history of evolution of the universe from the beginning to the end.  The results reveal an interesting correlation between Hubble's linear law and the initial and the final states of the universe.

\section{Strong cosmological principle}
 Based on the principle of least action involving the space-time metric $ds^2$, Hubble's law of motion for distant galaxies suggests a different space-time properties of the universe as a whole.  In a previous work,\cite{7} we consider the motion of distant galaxies in the super-macroscopic world with the usual cosmological principle related to homogeneity and isotropy of the universe.   With the help of Yang-Mills gravity, we obtain the Okubo equation of motion in an inertial frame,\cite{7} 
\be
G^{\mu\nu}(t)\p_\mu S \p_\nu S - m^2=0,
\ee
 where $G^{\mu\nu}(t)$ takes the diagonal form 
 \be 
 (B^{-2}(t), -A^{-2}(t),  -A^{-2}(t),  -A^{-2}(t)), \ \ \ \ \ \   B(t)=\b t^{1/2}, \ \ \ \ \   A(t)=\a  t^{1/2},
 \ee
  for the matter dominated universe in flat space-time.  The Okubo equation with the scale factors in (10) gives different upper limits for Hubble recession velocities associated with different galaxies in general, as measured in an inertial frame.  For comparison, the corresponding FLRW model based on general relativity gives the (Riemannian) metric tensor $g_{\mu\nu}(t) =(1, -a^{2}(t),  -a^{2}(t),  -a^{2}(t)), \ a(t) \propto t^{2/3}$.  The FLRW model does not have an upper limit for Hubble's recession velocity.  Since $g_{\mu\nu}(t)$ is not obtained in inertial frames, its comparison with $G_{\mu\nu}(t)$ is not completely satisfactory.

We simplify these properties to discuss  highly simplified dynamic models in particle-cosmology including quantum Yang-Mills gravity. We are interested in the correlation between Hubble's linear law\cite{8} and the beginning,\cite{9} the expansion and the future  of the universe.  For simplicity, a galaxy is idealized as a point-like object with a constant mass M.  All possible long-range forces in particle physics\footnote{They include the long-range forces produced by electric charges, baryon charges, lepton charges and masses.  There may be other sources.}  in the universe acting on this point-like galaxy are assumed to be included in an effective metric tensor $G_{\mu\nu}(t)$ of $ds^2$.  Of course, this is highly simplified.  This effective metric tensor in the Okubo action $S_{Ok}$ dictates that the motion a point-like galaxy follows the Okubo equation of motion  $G^{\mu\nu}(t)(\p_\mu S) \p_\nu S - m^2=0$.  The key postulate for the classical particle-cosmology is  the `strong cosmological principle':\footnote{It is suggested by Yang-Mills gravity, which leads to the above equation (10).} 
\bigskip

`{\em The effective metric tensor of the space-time for the universe as a whole takes the form  $G_{\mu\nu}(t)=\e_{\mu\nu}A^2(t) $ for all times $t \ge 0$ and all length scales $r \ge 0$ in an inertial frame $F=F(t,x,y,z)$.}'   
\bigskip

 Thus, the models based on strong cosmological principle have the following cosmic Okubo equation,
\be
A^{-2}(t)\e^{\mu\nu}(t)(\p_\mu S) \p_\nu S - m^2=0, \ \ \  t \ge 0, \ \ \   r \ge 0,
\ee
 in an inertial frame $F$, in which $\e^{\mu\nu}=(1.-1,-1,-1).$   
 
The advantages of a general space-time framework for inertial and non-inertial frames is that one has operationally defined space and time coordinates, so that one has a meaningful description of nature.  The operational meaning of space and time coordinate in non-inertial frames is more subtle.  Nevertheless, it can be realized physically. In an inertial frame, one can set up a grid of clocks to measure time and one has meter stick to measure length.  However, in a non-inertial system, there are local distortions of length and time so that the space-time coordinates must be operationally defined differently.  In this case, one sets up a system of computer devices, called `space-time clocks', in a non-inertial frame.  The space and time coordinates of these clocks are constantly self-adjusted by computers to satisfy the space-time coordinate transformations between this non-inertial frame and an inertial frame with a known function of the acceleration\cite{10}.   Such a space-time clock system can be set up if and only if the space and time coordinates transformations between non-inertial and inertial frames are known.  In this sense, the existence of inertial frame in the framework is crucial for the operationally defined space-time coordinates of general frames (inertial and non-inertial).  It seems that the mathematical framework of general relativity based on the principle of general coordinate invariance is too general and complicated to realize these operational  definitions of space and time coordinates for inertial and non-inertial frames in flat space-time.\cite{11}  Wigner wrote: ``Evidently, the usual statements about future positions of particles, as specified by their coordinates, are not meaningful statements in general relativity.  This is a point which cannot be emphasized strongly enough.....  It pervades all the general theory, and to some degree we mislead both our students and ourself when we calculate, for instance, the mercury perihelion motion without explaining how our coordinate system is fixed in space''.\cite{11}\footnote{Wigner also wrote ``Expressing our results in terms of the values of coordinates became a habit with us to such a degree that we adhere to this habit also in general relativity, where values of coordinates are not per se meaningful.}

\section{ Cosmic HHK model with a `bang'} 
 
 Based on particle-cosmology in flat space-time, the fundamental CPT invariance\footnote{The CPT invariance in particle physics\cite{1} implies exact lifetime and mass equalities between particles and antiparticles.  It also implies opposite electroweak- and chromo-interactions properties between particle and antiparticles, and exactly the same gravitational interaction between particles and antiparticles (in the sense of quantum Yang-Mills gravity in flat space-time).} suggests the Big Jets model \cite{12}, which assumes the creation of  two jets, which may be pictured as two fireballs moving away from each other following the conservation of linear momentum.  It assumes that somewhere in the vacuum there is enormous amount of energy converted into elementary particles such as quarks, antiquarks, gauge bosons, etc.  Each fireball has enormous amount of particles and antiparticles. After annihilations, collisions and decays, one fire ball happens to be dominated by protons, neutrons, electrons, etc.  The fire ball cool down and becomes the 3K matter blackbody that we observe today.  
  The CPT invariance implies that the other fire ball must be dominated by antiprotons, antineutrons, positron, etc. and that it will become a 3K antimatter blackbody. This resembles the phenomena of very high energy collisions of particles in laboratory.  However reasonable it may be, it is premature to claim that we understand what was going on in the beginning of the universe. 
  
 In the HHK model, we shall avoid the speculation of the creation of all elementary particles and the confinement of quarks to become baryons and mesons with the confining forces.  We simply assume that there are matter in the beginning  and that there are fundamental laws of motion etc. at time $t=0$.
 
 The basic assumptions of the HHK model\cite{12} based on the strong cosmological principle and Yang-Mills gravity are as follows:   We consider highly simplified motion of a galaxy, idealized as a mass point, in the universe dominated by matter\cite{7}.  The space-time of the whole universe is assumed to have an effectively metric tensor, which takes the diagonal form with four non-vanishing time-dependent $G_{\mu\nu}(t)$,  
 \be
   G_{\mu\nu}(t)= \e_{\mu\nu}A^{2}(t),  \ \ \ \ \   \e_{\mu\nu}=(1,-1,-1,-1),   \ \ \  t \ge 0,
 \ee
$$
ds^2=\e_{\mu\nu}A^2(t)dx^\mu dx^\nu, \ \  \ \ \ \ A=\a t^{1/2},  \ \ \ \  c=\hbar=1, 
$$
where the effective scale factor $A(t)$ expands as time increases.  These results follow from (10) and (11) in Yang-Mills gravity with the strong cosmological principle.

   The motion of a point-like galaxy with a mass M in the universe is, as usual, assumed to be determined by the cosmic Okubo action $S_{Ok}$ involving $ds^2$ in (2),  
 \be
S_{Ok}= - M \int ds = - M \int \sqrt{\e_{\mu\nu}A^{2}(t) dx^\mu dx^\nu}.
 \ee
  To derive the equation of motion from (13), we consider the covariant space-time variation $\de x^\mu$ with a fixed initial point and a variable end point, together with the actual trajectory.\cite{13,14} 

The action $S$ leads to a Hamilton-Jacobi type equation (11) of motion for a point-like distant galaxy with mass M, which was called cosmic Okubo equation for the motion of galaxies.  In isotropic universe, we could choose the coordinates $(t,r)$ instead of $(t,x,y,z)$ for the motion of distant galaxies.  Thus, the Okubo equation (11) takes the form,
\be
A^{-2}\left[(\p_t S)^2 - (\p_r S)^2\right] -M^2 = 0, \ \ \ \ \ \   r, t \ge 0, \ \ \ \ \ \   S=S_{Ok}.
\ee
Since $A(t)$ does not involve $r$,\cite{7} we have the conserved generalized momentum $p=\p_r S$. As usual, we look for an $S$ in the form\cite{14,7} $S=-f(t) + pr.$  The trajectory of a galaxy is determined by the equation $\p S/\p p= constant$. We obtained the following results, 
\be
\dot{r}=\frac{1}{\sqrt{1+m^2 A^2/p^2}}= \frac{1}{\sqrt{1+\Om^2 t}}, \ \ \ \   A= \a t^{1/2},
\ee
$$
r(t)=\frac{2}{\Om^2} \sqrt{1+ \Om^2 t}, \ \ \ \   \Om=\frac{m\a}{p},
$$
$$
p=\frac{ \ov{M} \dot{r}(t)}{\sqrt{1-\dot{r}^2}}=constant, \ \ \ \ \   \ov{M}=M A(t),
$$
in an inertial frame F.   The constant of integration constant in $r(t)$ is assumed to be negligible for lack of data.  The results in (14) and (15) can also be obtained by solving the Lagrange equation, where the Lagrangian, $L=-mA(t)\sqrt{1-\dot{r}^2}$ is given by  the action (13).

For cosmic redshift $z$ of light emitted from distant galaxies with non-constant recession velocity $\dot{r}$ in (15) can be derived similar to the accelerated Wu-Doppler effects, using accelerated Wu transformations.\cite{5,7}  The wave 4-vector $K_{e\mu}$ emitted from a distant galaxy is associated with an eikonal equation or the massless Okubo equation (14) with $m=0$ and $\p_\mu S \to \p_\mu \psi_e= k_{e\mu}$.  The wave 4-vector of light as measured in an inertial frame satisfies the usual eokonal equation $\et^{\mu\nu} \p_\mu \psi \p_\nu \psi = 0$ with $\p_\mu \psi= k_\mu$.  It is natural to treat the observed cosmic redshift based on the `covariant eikonal equation', $G^{\mu\nu}(t) \p_\mu \psi_e \p_\nu \psi_e = \et^{\mu\nu} \p_\mu \psi \p_\nu \psi $, which satisfies the principle of limiting continuation of physical laws.\cite{5}
The cosmic redshift z and acceleration $\ddot{r}(t)$ of this distant galaxy, together with  the deceleration parameter $q_0$,  can be derived,\cite{7}
\be
z=\frac{1+V_r}{\sqrt{1-V^2_r} }-1, \ \  or \ \ \  V_r\equiv \dot{r}=\frac{2z+z^2}{2+2z+z^2},
\ee
\be
\ddot{r}=  \frac{- \Om^2}{(1+ \Om^2 t)^{3/2}}, \ \ \ q_o =  \frac{- \ddot{A}(t_o) A(t_o)}{\dot{A}^2(t_o)} = 1,
\ee
 We note that $V_r=\dot{r}(t)$ in (16) is  the non-constant recession velocity of the distant galaxy at the moment when the light was emitted.  The time  $t_o$ is the present age of the universe.  Also, the value $q_o=1$ actually holds for all times because 
$$
q(t)= - {\ddot{A}(t) A(t)}/{\dot{A}^2(t)} = 1.
$$  
Nevertheless, the actually accelerations of distant galaxies is a function of time, as shown in $\ddot{r}$ in (17).  Thus, it appears that the acceleration $\ddot{r}=\ddot{r}(t)$ gives more physical information about the evolution of the universe than the deceleration parameter $q_o$ (or `function' $q(t)$) of the model.  

Equations in (15) lead to an exact inverse relation between recession velocity $\dot{r}(t)$ and distance $r(t)$, 
\be
\dot{r}(t) =\frac{2}{\Om^2} \frac{1}{r(t)},
\ee
which holds for all times.  However, one can also introduce `Hubble's function' $H(t)=\dot{A}(t)/A(t)$ to express (18) in the form,
\be
\dot{r}(t) \approx H(t) r(t), \ \ \ \    m>> p,
\ee
in non-relativistic approximations.  One usually takes $t=t_o$ as the age of the universe in $H(t)$  and call $H(t_o)$ the Hubble's constant\cite{11}.  If a set of data for (19) satisfies the approximation $H(t)\approx const.$ for $t \approx t_o$  then one could interpret (19) as the recession velocity of a galaxy is proportional to its distance for a certain time interval.  Nevertheless, when one looks at the whole history of the evolution, then the HHK model predicts that the recession velocity will slow down to zero in the future of the universe, as shown in (18).

At the first glance, it sounds strange that one has the relations (18) and (19) in the HHK model.  However, similar situation happened in the usual FLRW model based on general relativity, in which one has the corresponding  Hamilton-Jacobi type equation, 
\be
g^{\mu\nu}(t)(\p_\mu S)( \p _\nu S) - m^2 =0, \ \ \  g^{\mu\nu}=(1, -a^{-2},  -a^{-2},  -a^{-2}), \ \  a(t)=a_o t^{2/3},
\ee
 for matter dominated universe.  Follow the steps (13)-(19), the FLRW model gives the following relations,
\be
\dot{r}(t)  \approx \frac{H(t)r(t)}{2} \approx  \frac{9}{a^3_o r^2(t)}, 
\ee
$$
 \ \ H(t) =\frac{\dot{a}}{a}=\frac{2}{3t}, \ \ \ \ \ \  r(t) \approx \frac{ 3t^{1/3}}{a_o},
$$
where the approximation $m<< p$ is made for simplicity.  The relations in (21) resemble to those in (18) and (19).  These relations shall be further discussed, together with those of (18) and (19), in the last section.

  Thus, according to the exact solutions (15) and (17) of the cosmic Okubo equation (9) together with (10),  HHK model leads to the following picture for the initial and final states of the universe,  
\be
 r  = \frac{ 2}{\Om^2},  \ \    \dot{r} \to  1, \ \   \ddot{r}=  - \ \frac{ \Om^2}{2}, \ \  z\to  \infty, \ \ \ \   t \to 0;
\ee
\be
 r \to \infty, \ \ \    \dot{r}(t) \to 0, \ \ \ \   \ddot{r}(t) \to 0, \ \ \  z \to 0, \ \ \   t \to  \infty,
\ee
as observed from an inertial frame.  The cosmic redshift z is related to the recession velocity of a distant galaxy, as shown in (16).\cite{7}

\section{A model with strict liner relation for recession velocity and distance}

So far, cosmological model with known scale factors in the metric tensor give the Hubble's law in the time-dependent form
\be
\dot{r}(t) \approx H(t) r(t),
\ee
where one takes $H(t)=H(t_o)$ as the Hubble `constant', where $t_o \approx 14$ billion years.  As a result, it is usually interpreted that the recession velocity $\dot{r}$ in (18) is `proportional' to the distance of a distant galaxy.  The data of cosmic redshifts of distant galaxies are also interpreted to be in line with (18) with $H(t)=H(t_o)$.  However, if the data cover a really large time interval in the cosmic evolution, e.g., $\De t \approx 0.5 t_o$, the usual interpretation of $H(t)=H(t_o)$ in (18) may no longer be appropriate.  One may ask whether there is a model which implies
\be
\dot{r}(t) = const. \times r(t).
\ee
for a large time interval, if not for all times.  It is interesting to see what are the implications on the early universe.

Let us consider a model in flat space-time with inertial frames, in which the equation of motion for distant galaxies gives an approximate solution (19).  We shall assume a suitable time-dependent scale factors or metric tensors to satisfy the phenomenological property (19) and to investigate its physical implications.  For this purpose, we assume that the effective metric tensor $G_{\mu\nu}(t)$ are given by
\be
ds'^2 = A'^2 dt^2 - A'^2 dr^2,    \ \ \ \ A'=\a' t^ {-1/2 },
\ee
in a super-macroscopic space-time, in consistent with the cosmological principle.  As usual, it is natural to assume the action $S_{lr}$ associated with the linear relation ($lr$) (25) to be given by
\be
S_{lr}=- m \int ds'  \equiv \int L_{lr} dt, 
\ee
\be
L_{lr}= - m A'(t) \sqrt{1-\dot{r}^2}, \ \ \ 
\ee
where a distant galaxy is approximated by a point with a mass $m$.  Because of the homogeneity and isotropy in the super-macroscopic world, it suffices to consider the coordinates $(t,r)$ instead of (t,x,y,z) for the motion of distant galaxies.  Note that the Lagrangian $L_{lr}=L_{lr}(r, \dot{r}, t)$ in (22) is time-dependent and $r$ is cyclic.  The Lagrange equation leads to the conservation of the `generalized momentum' $p$ conjugate to $r$,
\be
p=\frac{\p L_{lr}}{\p \dot{r}} =\frac{mA'\dot{r}}{\sqrt{1-\dot{r}^2}} = const., \ \ \ \ \  \dot{r}=\frac{dr}{dt}.
\ee
It gives the recession velocity $\dot{r} = dr/dt$,
\be
\dot{r}=\frac{1}{\sqrt{1+m^2 A'^2/p^2}} =\frac{1}{\sqrt{1+\Om^{'2}/t}}, \
\ee
$$
=\frac{ t^{1/2}}{\Om'\sqrt{1+t/\Om^{'2}}}, \ \ \ \ \ \  \Om'=\frac{m\a'}{p},  \ \ \ \ \  A'= \a' t^{-1/2};
$$
From (30), we obtain the acceleration $\ddot{r}$,
\be
\ddot{r} = \frac{C'_o t^{-1/2}}{2\Om' (1+t/\Om^{'2})^{3/2}}.
\ee
After integration, the recession velocity (30) leads to a time-dependent radius $r(t)$,
$$
r(t)= \frac{1}{\Om'} \int\frac{t^{1/2} dt}{\sqrt{1+t/\Om^{'2}}}=2\Om^{'2}\int \frac{x^2 dx}{\sqrt{1+x^2}}
$$
$$
=2\Om^{'2}\left[ \frac{x}{2}\sqrt{x^2+1} - \frac{1}{2}ln(x+ \sqrt{x^2+1)}  \right] + r_o''
$$
\be
\approx  2\Om^{'2}\left[ \frac{ \ t^{1/2}}{2\Om'}\sqrt{\frac{t}{\Om^{'2}}+1} \ \right] ,
\ee
where $x=t^{1/2}/\Om'.$
The constant of integration constant $r_o''$ in (32) is assumed to be negligible.  The `deceleration function' of the linear-relation model is given by
\be
q'(t)= - {\ddot{A}'(t) A'(t)}/{\dot{A}^{'2}(t)} = -3.
\ee

Thus, the linear-relation model gives the following picture of the beginning and the future of the universe,
\be
r(t) =  0, \ \ \ \   \dot{r} = 0,  \ \ \ \ \   \ddot{r}  \to  \infty, \ \ \  for  \ \ \ \    t \to 0;
 \ee
 \be
r(t)  \to  \infty, \ \ \ \ \ \   \dot{r} \to 1, \ \ \ \  \ddot{r}  \to  0, \ \ \ for   \ \    t \to \infty.
\ee

\section{Discussions}

Let us consider the approximation $t/\Om'^2 << 1$ in the linear-relation model.  It  is the same as the non-relativistic approximation because
\be
 \frac{t}{\Om^{'2}}  \approx \frac{\dot{r}^2}{c^{2}} <<  1. 
\ee
We use natural units, $c=\hbar=1$. In this non-relativistic approximation, the linear-relation model  with the results (30) and (32) leads to the relation,
\be
\dot{r} \approx \frac{ t^{1/2}}{\Om'} \approx  \frac{r(t)}{  \Om^{'2} } ,  \ \ \ \ \ \ \ \ \   t/\Om' << 1,
\ee
where $ \Om'={m\b'}/{p}$ is a time-independent constant.   Thus, in the non-relativistic approximation, the linear-relation model gives  a `strict' linear relation between the recession velocity $\dot{r}$ and the distance $r(t)$ of a distant galaxy.

The `linear-relation' model is constructed on the basic assumption (26) and (27)  to understand the empirical Hubble's  law on the basis of a Lagrangian dynamics with the Lagrangian $L_{lr}$ in (28).    The result (37) suggests that we could use the experimental value of the Hubble constant $H_o$ in the linear Hubble's law,
\be
recession\ velocity \   \dot{r}(t) \ \ \approx \ \ H_o \times distance  \ r(t),
\ee
to estimate the unknown constant $\Om'$ in (24).   We obtain the result
\be
\Om^{'2} \approx  \frac{1}{H_o} \approx  9.778h^{-1} \times 10^9 years.
\ee
where $0.5 \le h  \le 0.85$.\cite{11} 

From (36) and (39), all the numerical  values of time-dependent functions such as $r(t')$, $\dot{r}(t')$ and $\ddot{r}(t')$ can be considered as non-relativistic approximation, provided $t' << 10^{10}$ years.

We would like to show possible correlations between Hubble's linear law and the beginning, the expansion and the future of the universe based on two highly simplified dynamic models.  These model are constructed on the basis of the principle of least action (involving $ds^2$), the strong cosmological principle and Yang-Mills gravity.  For convenience of comparison in the non-relativistic approximation $m>>p$, we summarize the relation for recession velocity and distance (as observed in an inertial frame) in each model as follows:

[A]  HHK model ($A(t)= \a t^{1/2}, H(t)=1/(2t)$)
$$
Velocity-distance  \ relation: 
$$
\be
  \dot{r}(t)  \approx H(t) r(t),  \ \ \ \ \ \ \ \ \    \dot{r}(t) \approx \frac{2}{\Om^2 r(t)}; 
\ee
$$
Initial (t=0) \ and \ final (t=\infty)\ states  \ of \ the \ universe: 
$$
$$
r(0)=\frac{2}{\Om^2}, \ \  \dot{r}(0) \to c; \ \ \ \ \ \ \ \ \ \ \ \  r(\infty)\to \infty, \ \  \dot{r}(\infty) \to 0.
$$

[B] Linear-relation model ($ A'(t) = a' t^{-1/2},\ \ H'(t)=-1/(2t)$)
$$
Velocity-distance  \ relation: 
$$
\be
 \dot{r}(t) \approx |H'(t)|\frac{2 r^3(t)}{ \Om'^4}, \ \ \ \ \ \ \ \ \    \dot{r}(t)  \approx \frac{r(t)}{\Om'^2};
\ee
$$
Initial (t=0) \ and \ final (t=\infty)  \ states  \ of \ the \ universe:  
$$
$$
r(0)= 0,  \ \  \dot{r}(0)= 0; \ \ \ \ \ \ \ \ \ \  r(\infty)\to \infty, \ \ \dot{r}(\infty) \to c,
$$

[C] FLRW model ($ a(t) = a_o t^{3/2},\ \ H(t)=\dot{a}/a(t)=2/(3t)$
$$
Velocity-distance  \ relation: 
$$
\be
 \dot{r}(t) \approx  \frac{H(t)r(t)}{2}, \ \ \ \ \ \ \ \ \    \dot{r}(t)  \approx \frac{9}{a_o^3 r^2(t)};
\ee
$$
Initial (t=0) \ and \ final (t=\infty)  \ states  \ of \ the \ universe:  
$$
$$
r(0)= 0,  \ \  \dot{r}(0)= \infty; \ \ \ \ \ \ \ \ \ \  r(\infty)\to \infty, \ \ \dot{r}(\infty) =0,
$$

  Thus, according to the exact solutions (15) and (17) of the cosmic Okubo equation (9) together with (10),  HHK model leads to the following picture for the initial and final states of the universe,  
\be
 r  = \frac{ 2}{\Om^2},  \ \    \dot{r} \to  1, \ \   \ddot{r}=  - \ \frac{ \Om^2}{2}, \ \  z\to  \infty, \ \ \ \   t \to 0;
\ee
\be
 r \to \infty, \ \ \    \dot{r}(t) \to 0, \ \ \ \   \ddot{r}(t) \to 0, \ \ \  z \to 0, \ \ \   t \to  \infty,
\ee
as observed from an inertial frame.  The cosmic redshift z is related to the recession velocity of a distant galaxy, as shown in (16).\cite{7} 

 According to the HHK model with the fundamental CPT invariance, if one assumes all particles in the universe now are created in the beginning of the universe.  The CPT invariance dictates that all interactions of particles must satisfy a maximum symmetry between particles and anti-particles.\cite{1}  The experimental observation that there are no equal amount of antiparticles in our observable portion of the universe suggests that the universe was created with two big jets rather than one big bang.  One may picture it as two gigantic fireballs flying away from each other in opposite directions.  Within each gigantic fireball, the processes of evolution will be similar to that of a big bang.
 
 The work was supported in part by Jing Shin Research Fund and Prof. Leung Memorial Fund of the UMassD Foundation.

\bigskip

\bibliographystyle{unsrt}

\begin{thebibliography}{99}

\bibitem{1}T. D. Lee, {\em Particle Physics and Introduction to Field Theory}(Harwood Academic Publishers, 1981); and ref. 14. 

\bibitem{2}J. P. Hsu, Int. J. Mod. Phys. {\bf 21}, 5119 (2006).

\bibitem{3}J. P. Hsu and L. Hsu, {\em Space-Time, Yang-Mills Gravity and Dynamics of Cosmic Expansion} (World Scientific, 2020) pp. 113-127.

\bibitem{4}J. P. Hsu,  Euro. Phys. J. Plus 126 (3), 24 (2011).

\bibitem{5}J. P. Hsu and L. Hsu, ref. 3. pp. 122-124.

\bibitem{6}F. Dyson, in {\em 100 Years of Gravity and Accelerated Frames:  The Deepest Insights of Einstein and Yang-Mills}. (Ed. J. P. Hsu and D. Fine, World Scientific, 2005) p. 348.

\bibitem{7}L. Hsu and J. P. Hsu, Chin. Phys. C {\bf 43} No. 10, 105103 (2019).

\bibitem{8}E. Hubble, Proc. Nat. Acad. Sci. (PNAS) {\bf 15}, 168  (1929).

\bibitem{9}A. Gamow, Nature {\bf 162} (4122) 680 (1948).

\bibitem{10}J. P. Hsu and L. Hsu, {\em A Broader View of Relativity, General Implications of Lorentz and Poincar\'e Invariance} (J. P. Hsu and L. Hsu, World Scientific, 2006) pp. 23-43.  

\bibitem{11}E. P. Wigner, {\em Symmetries and Reflections, Scientific Essays} (The MIT Press, 1967) pp. 52-53.

\bibitem{12}J. P. Hsu, Leon Hsu, D. Katz, Mod. Phys. Lett. A {\bf 33} 1850116 (2018).

\bibitem{13}J. P. Hsu and L. Hsu, ref. 3, pp. 223-228.

\bibitem{14}L. Landau and E. Lifshitz, {\em The Classical Theory of Fields} (Addison-Wesley, 1951) pp. 41-43, pp. 268-271, pp. 312-314.



\end{thebibliography}


\clearpage
\end{document}